# Generation of Ultrastable Microwaves via Optical Frequency Division

T. M. Fortier[*], M. S. Kirchner, F. Quinlan, J. Taylor, J. C. Bergquist,
T. Rosenband, N. Lemke, A. Ludlow, Y. Jiang, C. W. Oates and S. A. Diddams[*]

*National Institute of Standards and Technology*
*325 Broadway, Boulder, CO 80305 USA*
*email: tara.fortier@nist.gov, scott.diddams@nist.gov*

There has been increased interest in the use and manipulation of optical fields to address challenging problems that have traditionally been approached with microwave electronics. Some examples that benefit from the low transmission loss, agile modulation and large bandwidths accessible with coherent optical systems include signal distribution, arbitrary waveform generation, and novel imaging[1]. We extend these advantages to demonstrate a microwave generator based on a high-Q optical resonator and a frequency comb functioning as an optical-to-microwave divider. This provides a 10 GHz electrical signal with fractional frequency instability ≤8x10$^{-16}$ at 1 s, a value comparable to that produced by the best microwave oscillators, but without the need for cryogenic temperatures. Such a low-noise source can benefit radar systems[2], improve the bandwidth and resolution of communications and digital sampling systems[3], and be valuable for large baseline **interferometry[4], precision spectroscopy and the realization of atomic time[5-7].**

Several photonic systems, including optical delay-line oscillators[8], whispering-gallery-mode parametric oscillators[9] and dual-mode lasers[10] have been investigated for the generation of low noise microwave signals. An alternative approach, based on high quality factor ($Q$) optical resonator and all-optical frequency division, shows promise for the generation of microwaves with excellent frequency stability[6,7,11-13]. This is because low absorption and scattering in the optical domain can yield quality factors ($Q$) approaching $10^{11}$ in a room-temperature Fabry-Perot (FP) resonant cavity. For a well-isolated cavity, average fluctuations in the cavity length amount to ~100 attometers on a 1 s timescale. A continuous wave (CW) laser stabilized to such a cavity can reach fractional frequency instability as low as $\Delta\nu/\nu \sim 2\times10^{-16}$ for averaging times of 1-10 s [14-18]. Transfer of this stability to a microwave signal is the topic of this paper, where we demonstrate a 10 GHz electronic signal with exceptional frequency stability and spectral purity.

Figure 1 outlines the principle of the photonic oscillator we have developed. Phase coherent division of the stable optical signal to the microwave domain preserves the fractional frequency instability, while reducing the phase fluctuations by a factor of $\sim 5\times10^4 = $ (500 THz / 10 GHz). Such frequency division is accomplished by phase locking a self-referenced femtosecond laser frequency comb to the optical reference[11]. This transfers the frequency stability of the stable CW laser oscillator to the timing between pulses in the laser pulse train, and hence to a microwave frequency that is detected as the pulse repetition rate, ($f_r$ ~0.1-10 GHz). In the case of a high fidelity optical divider, the sub-Hertz optical linewidth of the reference laser is translated into a microhertz linewidth on $f_r$. A fast photodiode that detects the stabilized pulse train generates photocurrent at frequencies equal to $f_r$ and its harmonics, continuing up to the cutoff frequency of the photodiode.

Using this photonic oscillator approach, we demonstrate a 10 GHz signal with absolute instability ≤$8\times10^{-16}$ at 1 s of averaging. This corresponds to a phase noise of L(f) = -104 dBc/Hz at 1 Hz offset from the carrier, decreasing to near the photon shot-noise limited floor of -157 dBc/Hz at an offset of 1 MHz. The integrated timing jitter over this bandwidth is 760 attoseconds. This measurement represents a significant improvement over previous work, with a reduction of phase noise power by 10 to 1000 times across the measured spectrum (1 Hz - 1 MHz)[7] and a factor of 4 reduction in instability[7,11]. This absolute timing characterizes one of the lowest phase noise microwave signals generated by any source.

Since the microwaves generated from our photonic approach have a phase noise that is lower than that available from commercially available microwave references, characterization of the generated phase noise requires that we build and compare two similar, but fully independent systems. The optical dividers in our photonic systems are based on octave-spanning 1 GHz Ti:sapphire femtosecond lasers and cavity-stabilized lasers at 578 nm and 1070 nm ($\nu_{opt1}$ and $\nu_{opt2}$ in Fig. 1 and Fig. 2). Compared to 250 MHz Er:fiber combs[13], the 1 GHz Ti:sapphire combs provide a 25 dB reduction in the shot noise floor. Although the exact wavelength of the CW lasers is not critical for microwave generation, what is significant is that the two systems are independent. In fact the FP cavities are situated in laboratories on different wings and floors of our research building. The pulsed output of the frequency-stabilized Ti:sapphire laser illuminates a high-speed, fiber-coupled InGaAs P-I-N photodiode that produces a microwave signal at 1 GHz and harmonics up to ~15 GHz. A band pass filter selects the 10 GHz tone that is subsequently amplified in a low-phase noise amplifier. The amplified signal is combined on a mixer with a similar signal from the second system, and the output of the mixer is analyzed to determine the relative frequency and phase fluctuations. In addition to the 10 GHz microwave signal, we also measure the optical stability of the frequency comb and the CW lasers, thereby obtaining a lower limit of the timing stability of the microwave signals. This is accomplished by measuring and analyzing the optical beat signal $f_b$ between the second stabilized CW laser $\nu_{opt2}$ and a tooth of the frequency comb that is independently stabilized by $\nu_{opt1}$ (see Fig. 1).

Phase noise data are presented in Fig. 3. The absolute single-sideband phase noise L(f) on an individual 10 GHz signal is given by curve 3(a). This curve is 3 dB below the measured noise under the assumption that the contribution from both oscillators is equal and uncorrelated (see supplementary information). The phase noise from the optical heterodyne between the two CW lasers using one of the combs is given by curve 3(b), which has been normalized to the 10 GHz carrier. This represents the present noise floor given by a single

CW laser and the frequency comb. As can be seen, the optical and microwave data converge at -104 dBc/Hz at 1 Hz. Above 10 kHz, the noise floor is set by the photon shot noise of the 10 GHz photodetector. Curve 3(c) shows the calculated shot noise floor of -157 dBc/Hz for the 10 GHz signal delivered at a power level of -8 dBm from 4 mA of average photocurrent. In the range of 10 Hz to 1 kHz, the noise contribution of microwave amplifiers cannot be neglected, as shown in curve 3(d). The combined noise of the CW laser, frequency comb, amplifiers and shot noise is given by curve 3(e). There is good agreement between this projection and the actual measurement, indicating that we have identified and properly accounted for the present limitations to the noise floor.

The spurious peaks in the 10 GHz phase noise (Fig. 3(a)) between 5 Hz and 300 Hz arise from unidentified intermittent noise sources that also appear on the optical comparison. The largest spur at 29 Hz is a known vibration of our laboratory floor. The microwave data in Fig. 3(a) was chosen to show the upper limit to the phase noise. Optical data without the spurs (trace 3(b)) was chosen to display the lower limit to the phase noise with our current optical references and optical dividers, neglecting limitations due to photodetection of $f_r$. The right axis of Fig. 3 shows that even the largest spurs are sub-femtosecond, and the integration over 1 Hz to 1 MHz yields a timing jitter of 760 attoseconds. The extension of this integration to 5 GHz at the present shot noise level yields timing jitter of ~25 fs. Straightforward reduction of the noise floor with band-pass filters provides still lower integrated jitter.

In Figure 4, the corresponding frequency counter data show the instability of the 10 GHz microwave signals and the optical instability of the CW lasers and frequency comb. The time record of frequency counter measurements (1 s gate time) is shown in Fig. 4(a) and the fractional frequency instability calculated from these data are in Fig. 4(b). Under the assumption of equal and uncorrelated oscillators, the data of Fig. 4(b) have been reduced by a factor of √2 from the measurement. We have not post-processed these data, and the slow oscillations and linear drift seen in Fig 4(a) are the result of temperature variations of the independent FP cavity references.

The close-to-carrier phase noise and short-term instability with our photonic approach are lower than that achieved with any other room-temperature 10 GHz oscillator. With a thermal-noise-floor-limited optical cavity, a phase noise of L(f) -117 dBc/Hz at a 1 Hz offset appears feasible [13,18]. Even lower phase noise levels could be achieved in the future with new optical references based on spectral hole burning techniques [19]. As seen in Figure 5, the present noise is comparable to only the very best cryogenic dielectric oscillators [20-23]. Fiber delay-line oscillators have achieved lower noise floors at Fourier frequencies >1 kHz[8], but all such photonic devices have a noise floor ultimately limited by shot noise and the power handling capabilities of the high-speed photodiode (see supplementary information). In our case, a higher repetition rate comb would alleviate photodiode saturation effects, and a noise floor at high frequency near -165 dBc/Hz appears achievable[24]. A still lower noise floor would require higher-power photodetectors or a hybrid approach with a low-noise dielectric sapphire oscillator[25,26] locked to our photonic oscillator with bandwidth of ~1 kHz.

*Methods:*

**Optical reference oscillators:** Although details pertaining to the 518 THz [18] and 282 THz [14] CW reference lasers differ, we offer a general description of the systems here. Both oscillators are based on fiber and solid-state lasers that are frequency stabilized to a single transverse and longitudinal mode of a high finesse optical cavity via the Pound-Drever-Hall locking scheme. The cavities are constructed of low expansion ULE spacers with optically contacted high-reflectivity mirrors that exhibit a finesse of 200,000 and 300,000 for the 518 THz laser and the first harmonic of the 282 THz laser, respectively. In both systems, the intensity of the light incident on the high finesse cavities is stabilized to minimize thermal instabilities of the cavity length due to heating of the mirrors. Mounting of the cavities and the cavity geometries themselves, although different, are both chosen to minimize the effects of accelerations on the optical cavity length. The design of the FP cavities for the 518 THz and 282 THz FP cavities are similar to those described in references [15,16] and [14], respectively. To isolate the cavities from external perturbations, each cavity is held in a temperature-controlled evacuated chamber that is mounted on an active vibration stage inside an acoustic isolation enclosure. The light generated from the two systems have demonstrated optical linewidths < 1 Hz and a frequency instability that is < $7\times10^{-16}$ at 1 s of averaging. For historical reasons, the two optical reference cavities are separated by ~300 m, being located in labs on different floors of the NIST laboratory building. The frequency-stabilized light from the optical cavities is transmitted (with negligible change in optical stability or phase noise) via stabilized fiber optical links[27] of 30- and 300-m length to the optical frequency combs which are located in a third laboratory. We note that recent efforts have focused on characterizing and minimizing the acceleration sensitivity of such high-stability optical reference oscillators, including field tests where active cancellation of acceleration-induced frequency drifts have been demonstrated (see supplementary information).

**Frequency comb details and stabilization:** The optical frequency combs are Ti:sapphire ring lasers with a cavity length L=30 cm, which gives a repetition rate of 1 GHz, or a spacing between adjacent pulses of 1 ns. One laser system sits on a passively isolated (air legs) optical table and is enclosed in nested aluminium and plexiglass boxes. The second comb system is separated from the first by a few meters. It is enclosed in a free-standing isolation box that provides ~30 dB of acoustic suppression. The base plate of this comb is isolated from seismic vibrations by a piezo-actuated platform. Each laser is pumped with ~8 W of 532 nm light, and produces ~1W of modelocked power. Both lasers produce an optical spectrum with usable bandwidth from 550 nm to 1200 nm, which exceeds the gain bandwidth of the laser [28]. As a result, measurement of the offset frequency ($f_o$) is obtained directly from the laser by doubling the low frequency end of the laser spectrum at 1100 nm and referencing it to the high frequency end at 550 nm[28]. Optical interference of these two signals provides $f_o$, which is filtered, amplified and mixed with an rf reference frequency ($f_{rf}$) to provide an error signal that is fed back to the laser pump power to maintain the condition $f_o = f_{rf}$. A heterodyne beat between the CW laser frequency ($\nu_{opt}$) and a single mode of the self-referenced frequency comb provides an error signal, which is employed in an active servo loop that controls the cavity length using a piezoelectric actuated mirror to force the comb mode to oscillate in phase with $\nu_{opt}$ [28]. While the Ti:sapphire systems described here demonstrate excellent performance in the laboratory, more robust Er:fiber frequency combs have been demonstrated that have excellent close-to-carrier phase noise performance[13] and low acceleration sensitivity. Such systems could be an important component for a future optical frequency divider that operates outside the laboratory (see supplementary information).

**Photodetection and measurement system:** Photodetection of the laser repetition rate is accomplished using a pair of jointly-packaged, fiber coupled, 12-GHz, InGaAs P-I-N photodiodes (50 Ω terminated, + 9 V bias). The photodiodes are 60 μm in diameter, have responsivity of 0.34 A/W at 900 nm, and feature a 0.3 μm InP cap layer. Previously identified limitations are addressed by the combination of larger detector area and a thinner InP cap layer, which improve the power handling and reduce the conversion of amplitude noise to phase noise (AM-to-PM) in photodetection[29]. Light near 980 nm (with ~50 nm bandwidth) is coupled to the photodiodes via a 5 m fiber optical cable from one comb system and a 2 m fiber optical cable from the second comb system. The residual intensity noise (RIN) on this light is near -100 dBc/Hz at 1 Hz offset, which we estimate to not significantly impact the present phase noise (via AM-to-PM in the photodiodes). With ~12 mW of light incident on the photodiodes, we directly obtain approximately -8 dBm in the 10 GHz carriers which is then amplified to between 0 dBm and +7dBm for input to the mixer. At Fourier frequencies above ~10 kHz the achieved phase noise of Fig. 3 approaches the shot-noise limited floor of -157 dBc/Hz. In the absence of photodiode saturation, this phase noise floor should decrease proportionately to the detected optical power, implying a 10-fold increase in optical power leads to a 10 dB decrease in the noise floor (see supplementary information).

For phase noise measurements, the repetition rates of the two combs are adjusted such that the beat between the two 10 GHz signals is ~1-5 MHz. This mixed-down signal is input to a digital phase-noise measurement system that employs cross spectrum analysis to reduce the white noise floor below -160 dBc/Hz (depending on duration of averaging). For the counting measurements, the offset beat between the two 10 GHz signals is tuned to be ~50 kHz. The output of the mixer is low-pass filtered, amplified and input to a high-resolution Λ-type counter [30]. The fractional frequency instability of Fig. 4(b) is calculated from a time series of these counter measurements for both the microwave and optical data. By integration of the appropriately-weighted phase noise spectrum of Fig. 3, we verify that the 1 s instability presented here is consistent with the counter data of Fig. 4 and with the more conventional Allan Deviation[30]. Further details about microwave phase noise measurements at the low levels described here can be found in Refs. [8,20-23,25].

*Acknowledgements:* We thank A. Hati, L. Hollberg, D. Howe, C. Nelson, N. Newbury, and S. Papp for their contributions and comments on this manuscript, and A. Joshi and S. Datta of Discovery Semiconductor for providing the 10 GHz InGaAs photodiodes. This work was supported by NIST. It is a contribution of an agency of the US government and is not subject to copyright in the US.

*Author contributions:* T.M.F, M.S.K, F.Q, J.T and S.A.D. built, characterized, and operated the femtosecond lasers and measurement systems. J.C.B, T.R, N.L., A.L., Y.J., and C.W.O. constructed and operated the stable CW laser sources. S.A.D, T.M.F., and F.Q. acquired and analyzed the data and prepared the manuscript.

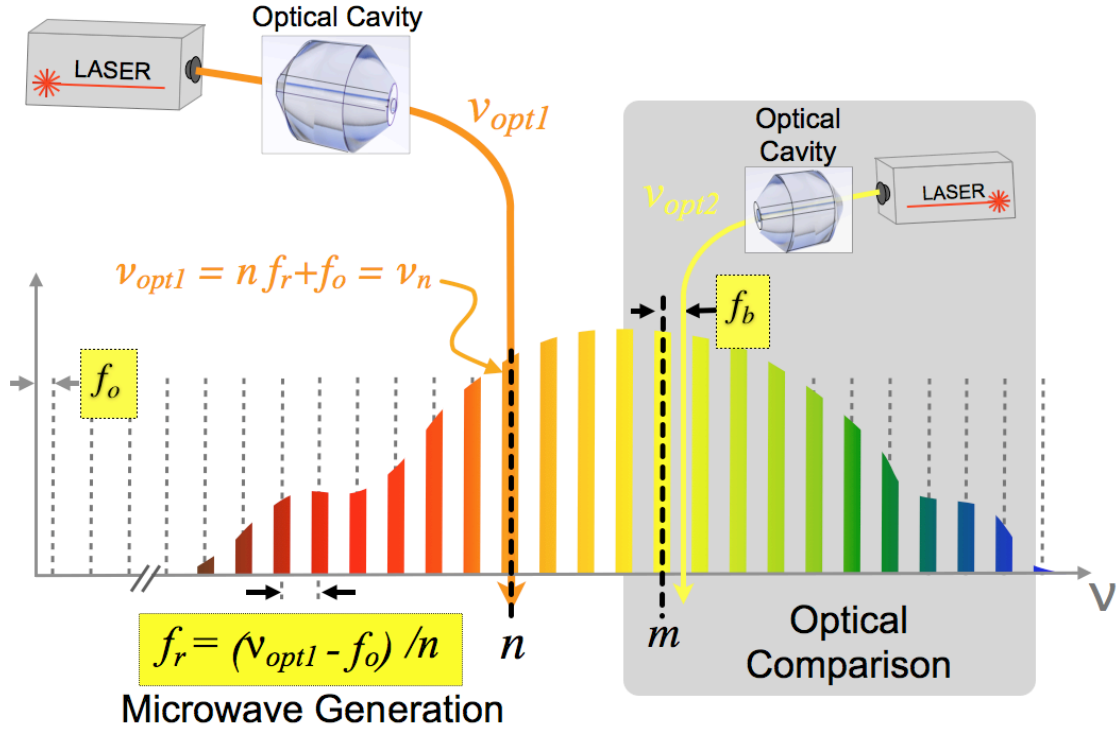

Figure 1: Depiction of how the optical frequency comb (OFC) from a mode-locked laser can be used as an optical frequency divider and as an optical frequency comparator. The OFC spectrum is stabilized by phase locking the $n^{th}$ comb element to $v_{opt1}$ (the frequency of a CW laser stabilized to a passive optical cavity), while simultaneously stabilizing of the laser offset frequency $f_o$ to a radio frequency reference. Pulse formation via passive mode-locking enforces a constant phase among the laser modes. As a result, stabilizing one mode to $v_{opt1}$ will transfer the optical cavity's stability to every optical mode of the OFC, as well as to the mode spacing $f_r$. The optical frequency is divided such that $f_r = (v_{opt1} - f_o)/n$. For the case where $f_o=0$, the beat signal $f_b = v_{opt2} - (m/n) v_{opt1}$ between a second stabilized CW laser ($v_{opt2}$) and mode $m$ of the OFC provides a measurement of the relative stability of $v_{opt1}$ and $v_{opt2}$ plus any excess frequency noise from the comb stabilization.

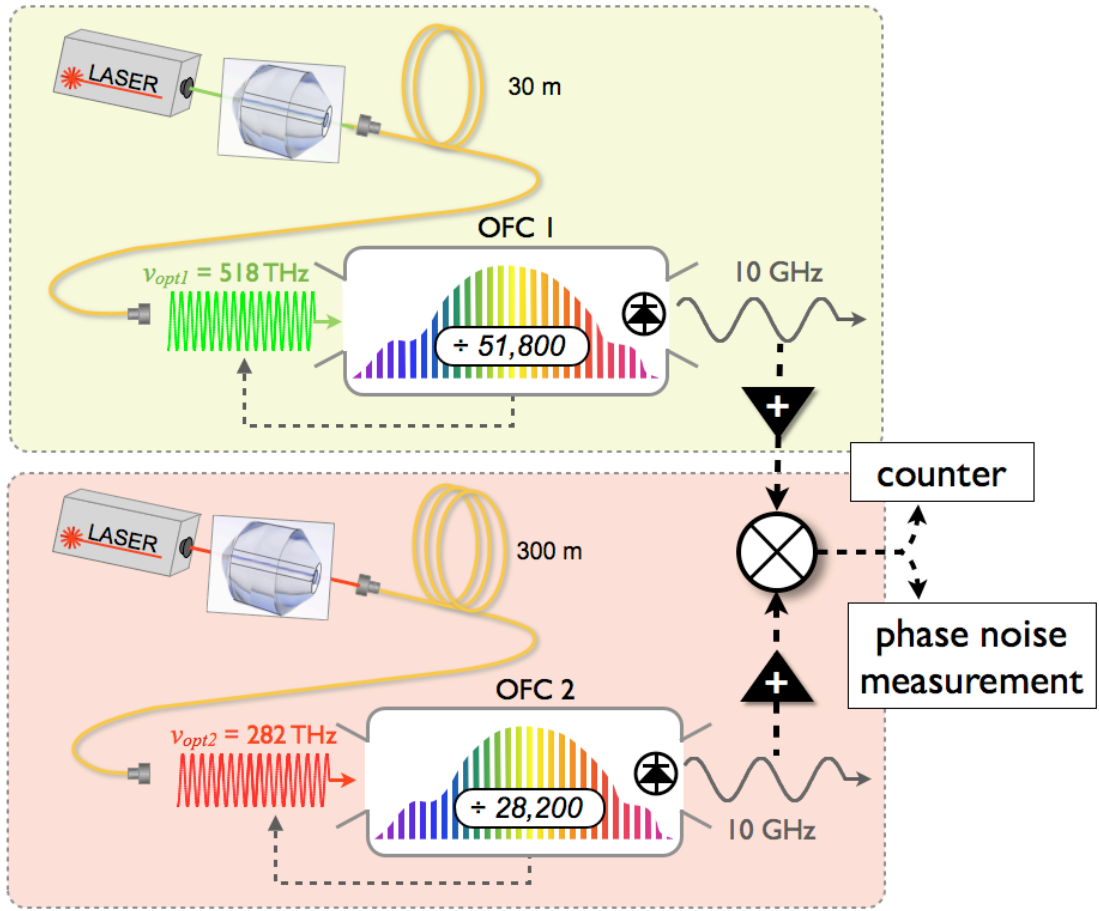

Figure 2: Schematic of the experimental setup used for generation and characterization of the 10 GHz low-noise microwaves. In each independent system, an OFC based on a 1 GHz Ti:sapphire mode-locked laser is phase locked to the optical frequency of CW laser that is stabilized to a passive optical cavity. The two cavities are located in different parts of the NIST laboratory building and the stable light is transferred to the OFC's via optical fiber. The 10th harmonic of the photodetected repetition rate yields a 10 GHz microwave signal that is phase coherent with $\nu_{\text{opt},i}$. The 10 GHz microwave signals generated from each system are filtered and amplified, and the mixed down product is characterized via frequency and phase noise measurements.

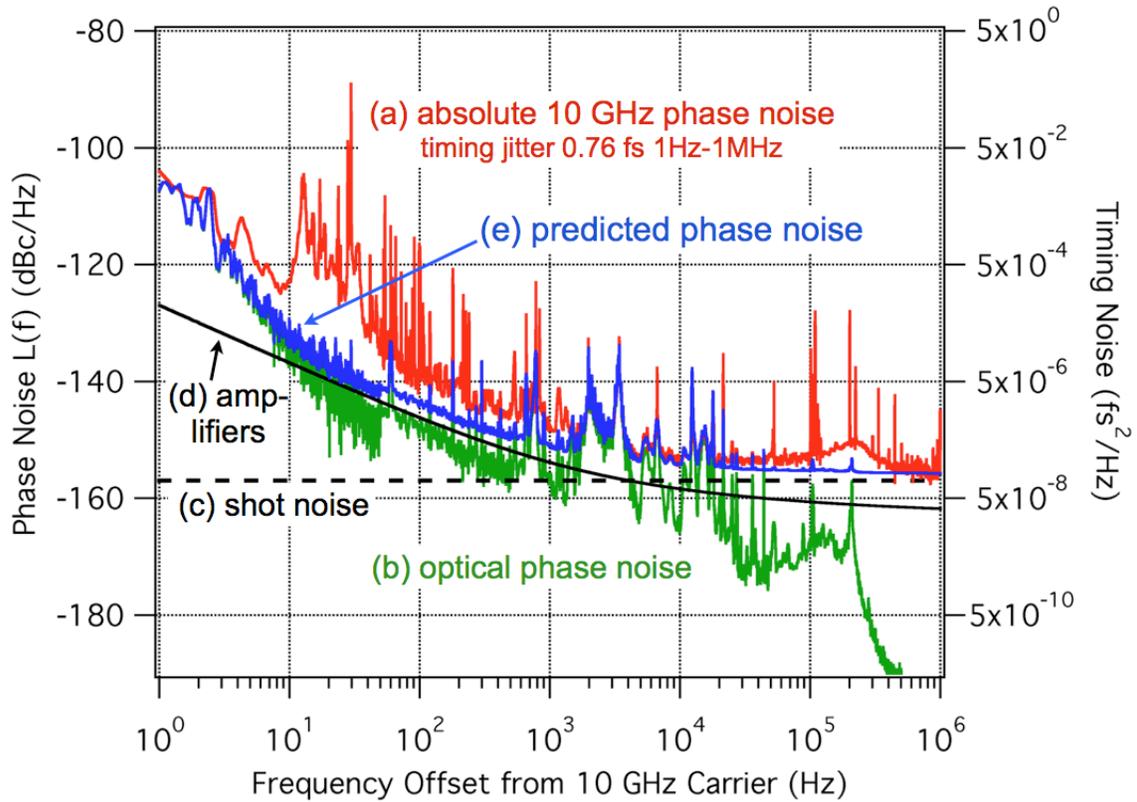

Figure 3: (a) Red trace – measured phase noise relative to the 10 GHz carrier for a single photonic oscillator. (b) Green trace – measured phase noise of a single optical reference (scaled to 10 GHz) as determined from $f_b$ of Fig. 1. This trace is the comparison of two optical oscillators with a single optical frequency comb, and it represents the present microwave phase noise floor if photodetection of the comb repetition rate were flawless (no amplitude-to-phase conversion, no shot noise limitations and no electronic noise). (c) Dotted black trace – calculated shot noise floor for 4 mA of average photocurrent generated via photodetection of the laser repetition rate. (d) Solid black trace - specified amplifier noise floor (extrapolated below 10 Hz). (e) Blue trace – sum of the dotted black, solid black and green traces, yielding the estimated phase noise achievable with the current systems.

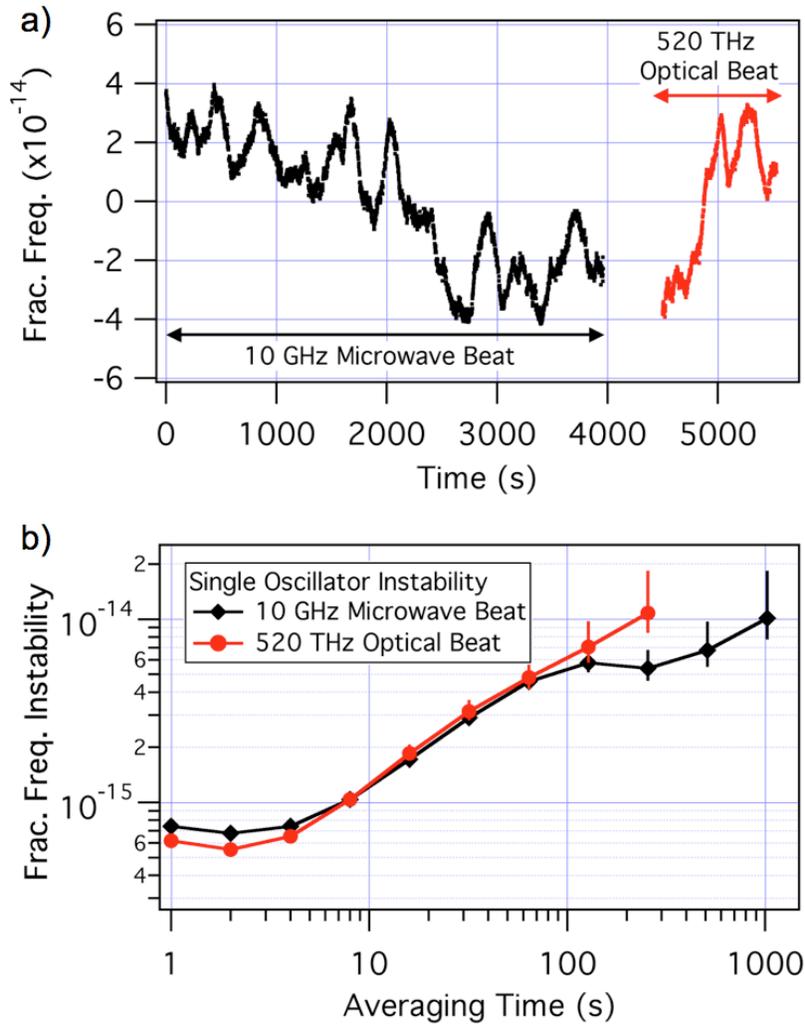

Figure 4: a) Time record of the measured beat frequency between the two photonically generated 10 GHz signals. Also shown is the beat signal ($f_b$ of Fig. 1) of the optical comparison of the two cavity stabilized reference lasers via a single frequency comb. b) Fractional frequency instability, calculated from the data of Fig. 4a) for a single oscillator assuming equal contributions to the instability from each oscillator used in the 10 GHz microwave and optical comparisons. The slightly higher instability for the 10 GHz signals at 1s is attributed to spurious peaks close to carrier from 10 to 100 Hz as seen in Figure 3.

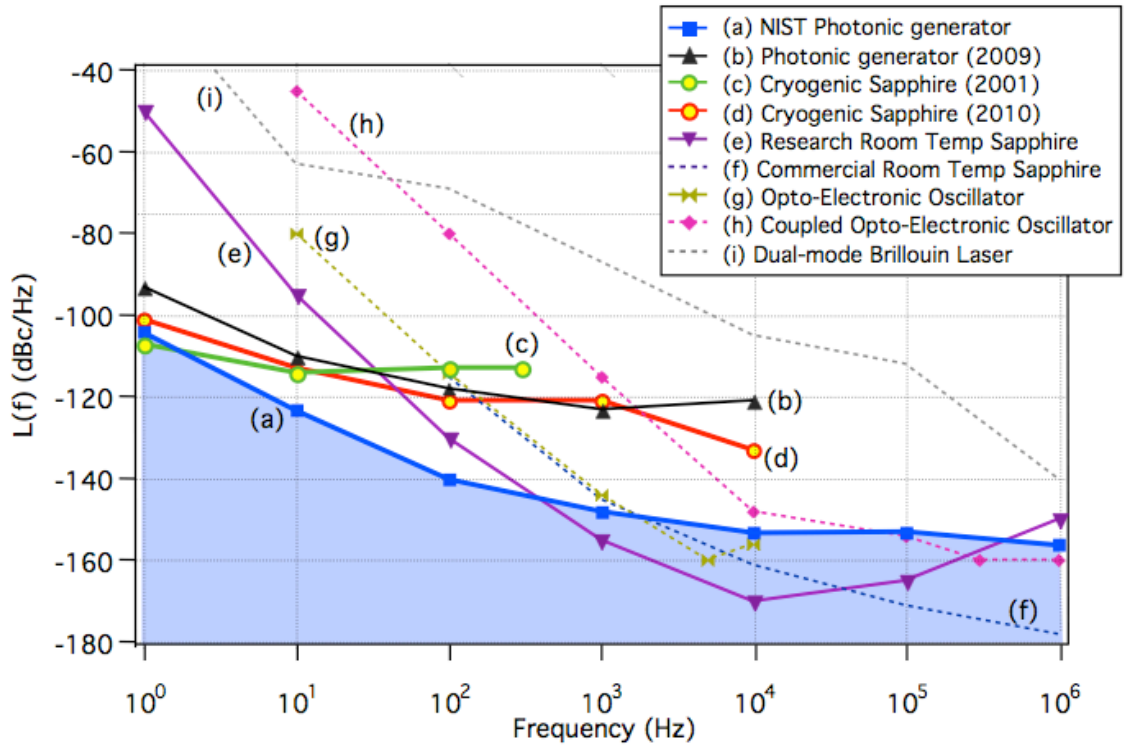

Figure 5. Approximate single sideband phase noise for several of the leading microwave generation technologies in the 10 GHz range. Spurious tones have been neglected for all data. (a) Results of this present work; (b) Previous Er:fiber and Ti:sapphire optical frequency divider results [7]; (c,d) Cryogenic sapphire oscillators [20,23] ; (e) Research room temperature sapphire oscillator [25] ; (f) Commercial room temperature sapphire oscillator [26]; (g) Long-fiber and (h) coupled opto-electronic oscillators [8]; (i) Dual-mode Brillouin laser [10].

# Supplementary Information for "Generation of Ultrastable Microwaves via Optical Frequency Division"

T. M. Fortier[*], M. S. Kirchner, F. Quinlan, J. Taylor, J. C. Bergquist,
T. Rosenband, N. Lemke, A. Ludlow, Y. Jiang, C. W. Oates and S. A. Diddams[*]

*National Institute of Standards and Technology*
*325 Broadway, Boulder, CO 80305 USA*
email: tara.fortier@nist.gov, scott.diddams@nist.gov

This supplement provides additional details on the impact of the femtosecond laser optical frequency dividers on the 10 GHz phase noise, and discusses properties (acceleration sensitivity) of stable laser oscillators and femtosecond laser frequency combs that will be relevant for systems use in non-laboratory applications.

**1) Residual phase noise of the optical frequency dividers and its effect on the generated microwave phase noise**

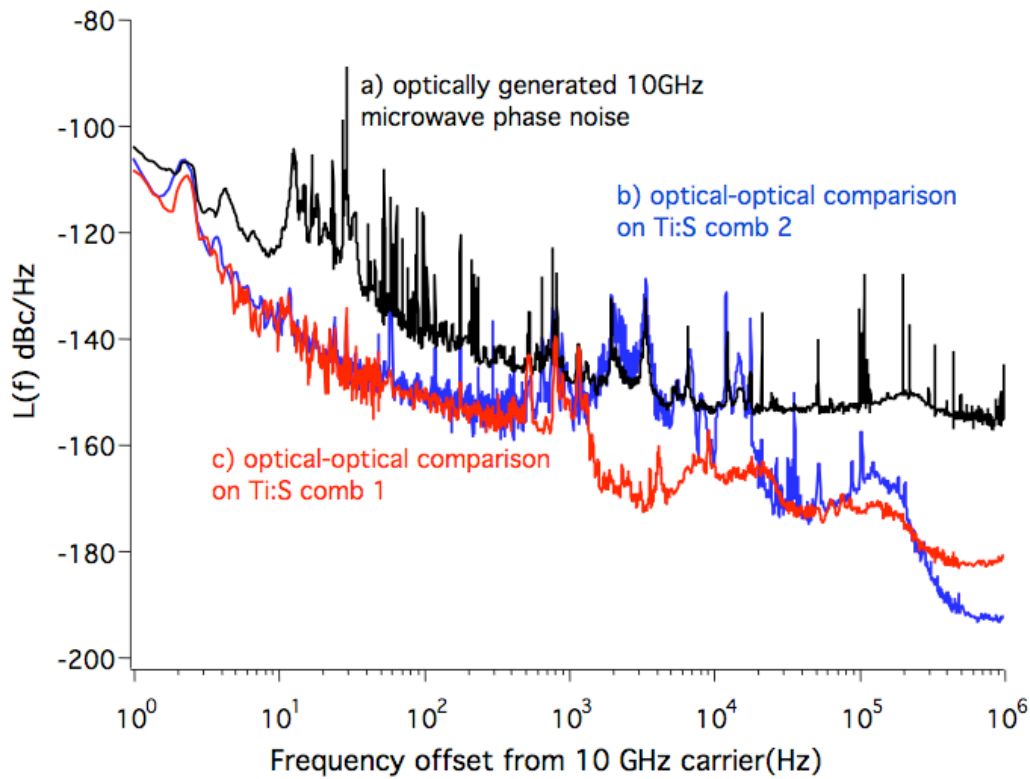

**Figure S1.** Phase noise of the optical-optical comparisons of the 282 THz and 518 THz reference lasers as measured on Ti:sapphire (Ti:S) optical frequency comb dividers 1 and 2. The data are scaled to 10 GHz assuming perfect division by

518,000.  Also shown for reference is the phase noise of the 10 GHz microwave-comparison signal obtained using the same optical dividers.

The phase noise of the optical-to-optical comparisons of the 282 THz and 518 THz reference lasers as measured on both the optical frequency comb dividers used in this work is shown in Fig. S1. The difference in the phase noise between in the curves in Fig. S1 b) and S1 c) shows the influence of the residual noise of the dividers on the laser repetition rate.  The main conclusion from these data is that only in a limited range of 1 kHz to 20 kHz does the optical noise of one of the frequency combs (Ti:S comb 2) significantly influence the 10 GHz microwave phase noise). Separate measurements of the optical divider residual noise shows no impact to the optical phase noise for offset frequencies < 1 kHz.  Because the residual noise of the optical dividers is low (~ -170 dBc/Hz), close to carrier (< 1 kHz) noise on the microwave signal is dominated by noise from the optical references, optical photodetection and measurement electronics. The low phase noise of -104 dBc/Hz at 1 Hz offset reported in our manuscript is the result of the extremely low relative drift of the optical cavities, which is under 0.1 Hz/s at the optical frequencies.

**2) Potential for a field-able low noise microwave generator based on optical frequency division:**

Although the data presented in our manuscript were measured in a laboratory environment, vibration sensitivity and overall size will become important parameters for the realization of a deployable or commercial microwave generator.   The main point to convey here is that the vibration sensitivity of some of the main components of our photonic oscillator approach have been measured, and values are competitive with, or even better than that achieved with more conventional microwave oscillators.

The development of transportable optical reference cavities [S1] and transportable optical frequency comb dividers [S2] is currently being pursued.  For instance, Ref. [S2] demonstrates the acceleration sensitivity of a fiber-laser-based femtosecond laser frequency comb below $2 \cdot 10^{-12}$/g for modulation frequencies < 3kHz.   In addition, Ref. [S1] demonstrates research on spherical high-finesse optical cavities with passive vibration sensitivity measured to be less than $\sim 3 \cdot 10^{-10}$/g.  Approaches to measure and actively compensate the effects of vibration-induced phase noise on an optical frequency reference have been demonstrated and reported by Thorpe et al. [S3].  These ideas have been extended to high-finesse optical cavities operating outside the laboratory to achieve acceleration sensitivity below $10^{-11}$/g [S4].  Significantly, we note that these acceleration sensitivities are comparable or better than those reported for the best room temperature and cryogenic sapphire oscillators, which are in the range of $1\text{-}3 \cdot 10^{-10}$/g [S5,S6].

While the system we employ uses different reference cavities and frequency combs than those described in Ref. [S1] and Ref. [S2], we don't see fundamental limitations that would restrict the approach we present from being made smaller and more robust against vibrations.  Indeed, [S1-S4] illustrate that realistic implementations already exist, and could likely be improved, to enable the operation in non-laboratory settings. For example, the combination of the small spherical cavities described in [S1] and the all-fiber frequency comb technology described in [S2] could form the core of a significantly more compact system, occupying just a few square feet of table space.

Finally, we note that the low-loss transmission of optical fiber provides the possibility to distribute the various components of our system over significant distances, which could have an advantage in some applications. This could enable the most environmentally-sensitive elements to be located in a more benign environment.

**3) Routes to reduction of the shot-noise limited floor**

In our work, we have achieved a shot-noise limited phase noise floor of -157 dBc/Hz. Assuming no photodiode saturation, the shot noise limited signal-to-noise ratio improves linearly with optical power. Straightforward scaling our present results with the same photodiode responsivity would imply that 100x more power (roughly 1.2 W of optical power and 400 mA of photocurrent) is needed to approach the state-of-the-art level of -180 dBc/Hz (1 MHz offset from 10 GHz carrier). This is a significant amount of optical power and attaining such low levels would be very challenging. However, there are several approaches being pursued in the active field of high-power photodiodes that could potentially lead to the attainment of this goal.

High-power, high-linearity photodiodes already exist that can provide ~23 dBm of power (10-30 GHz) with average photocurrents of 100-200 mA [S7]. Combining the microwave power from an array of independently-illuminated diodes is a viable approach that has also been demonstrated [S8]. Additionally, our present photodiodes are an InGaAs structure with relatively low responsivity of 0.35 A/W at 980 nm. Photodiodes with 2-2.5x that responsivity are available in the 1550 nm spectral region. This improvement alone would reduce the required optical power to ~500 mW, which could potentially illuminate an array of 4 photodiodes providing ~100 mA each and a combined 10 GHz power near 30 dBm.